\documentclass[a4paper,11pt]{article}

\usepackage[nodayofweek]{datetime}

\usepackage{enumitem}
\usepackage{float}
\usepackage[margin=2.5cm]{geometry}
\usepackage{graphicx}
\usepackage[labelfont=bf,labelsep=period]{caption}
\usepackage{times}
\usepackage{url}
\usepackage{xspace}
\usepackage{numbertabbing}

\usepackage{amsmath} 
\allowdisplaybreaks[2]          

\usepackage{amssymb} 

\usepackage{amsthm} 

\usepackage[usenames,dvipsnames]{xcolor} 
\usepackage{centernot} 
\usepackage[normalem]{ulem} 

\usepackage{hyperref}

\newcommand\myshade{85}
\colorlet{mylinkcolor}{Maroon}
\colorlet{mycitecolor}{RoyalBlue}
\colorlet{myurlcolor}{Aquamarine}

\hypersetup{
  linkcolor  = mylinkcolor!\myshade!black,
  citecolor  = mycitecolor!\myshade!black,
  urlcolor   = myurlcolor!\myshade!black,
  colorlinks = true,
}

\newtheoremstyle{plain-boldhead}
  {\topsep}
  {\topsep}
  {\itshape}
  {}
  {\bfseries}
  {.}
  { }
  {\thmname{#1}\thmnumber{ #2}\thmnote{ (\bfseries #3)}}
\newtheoremstyle{definition-boldhead}
  {\topsep}
  {\topsep}
  {\normalfont}
  {}
  {\bfseries}
  {.}
  { }
  {\thmname{#1}\thmnumber{ #2}\thmnote{ (\bfseries #3)}}
\theoremstyle{plain-boldhead}
\newtheorem{theorem}{Theorem}

\newtheorem{lemma}[theorem]{Lemma}

\theoremstyle{definition-boldhead}
\newtheorem{definition}[theorem]{Definition}
\newtheorem{remark}[theorem]{Remark}
\newtheorem{example}[theorem]{Example}

\newtheorem{construction}[theorem]{Construction}

\floatstyle{ruled}
\newfloat{algo}{tbhp}{algo}
\floatname{algo}{Algorithm}

\newcommand{\BF}{\ensuremath{\mathbb{F}}\xspace}

\newcommand{\BQ}{\ensuremath{\mathbb{Q}}\xspace}

\newcommand{\CA}{\ensuremath{\mathcal{A}}\xspace}
\newcommand{\CB}{\ensuremath{\mathcal{B}}\xspace}

\newcommand{\CF}{\ensuremath{\mathcal{F}}\xspace}
\newcommand{\CG}{\ensuremath{\mathcal{G}}\xspace}
\newcommand{\CH}{\ensuremath{\mathcal{H}}\xspace}

\newcommand{\CP}{\ensuremath{\mathcal{P}}\xspace}
\newcommand{\CQ}{\ensuremath{\mathcal{Q}}\xspace}
\newcommand{\CR}{\ensuremath{\mathcal{R}}\xspace}
\newcommand{\CS}{\ensuremath{\mathcal{S}}\xspace}
\newcommand{\CT}{\ensuremath{\mathcal{T}}\xspace}

\newcommand{\CW}{\ensuremath{\mathcal{W}}\xspace}
\newcommand{\CX}{\ensuremath{\mathcal{X}}\xspace}

\providecommand{\naive}{na\"{i}ve\xspace}
\providecommand{\Naive}{Na\"{i}ve\xspace}

\def \ifempty#1{\def\temp{#1} \ifx\temp\empty }

\newdateformat{simple}{\THEDAY\ \monthname[\THEMONTH]\ \THEYEAR}
\simple

\begin{document}

\title{\bf How to Trust Strangers:\\ Composition of Byzantine Quorum Systems}

 \author{Orestis Alpos \\ 
   University of Bern \\
   \url{orestis.alpos@inf.unibe.ch} 
   \and Christian Cachin\\
   University of Bern\\
   \url{cachin@inf.unibe.ch} 
      \and Luca Zanolini\\
   University of Bern\\
   \url{luca.zanolini@inf.unibe.ch}
 }

\date{}

\maketitle

\begin{abstract}\noindent
Trust is the basis of any distributed, fault-tolerant, or secure system. A \emph{trust assumption} specifies the failures that a system, such as a blockchain network, can tolerate and determines the conditions under which it operates correctly. In systems subject to Byzantine faults, the trust assumption is usually specified through sets of processes that may fail together. Trust has traditionally been \emph{symmetric}, such that all processes in the system adhere to the same, global assumption about potential faults. Recently, \emph{asymmetric} trust models have also been considered, especially in the context of blockchains, where every participant is free to choose who to trust.

  In both cases, it is an open question how to compose trust assumptions. Consider two or more systems, run by different and possibly disjoint sets of participants, with different assumptions about faults: how can they work together? This work answers this question for the first time and offers composition rules for symmetric and for asymmetric quorum systems. These rules are static and do not require interaction or agreement on the new trust assumption among the participants. Moreover, they ensure that if the original systems allow for running a particular protocol (guaranteeing consistency and availability), then so will the joint system. At the same time, the composed system tolerates as many faults as possible, subject to the underlying consistency and availability properties.

  Reaching consensus with asymmetric trust in the model of personal Byzantine quorum systems (Losa~\emph{et al.}, DISC 2019) was shown to be impossible, if the trust assumptions of the processes diverge from each other.  With asymmetric quorum systems, and by applying our composition rule, we show how consensus is actually possible, even with the combination of disjoint sets of processes.
\end{abstract}

\section{Introduction}

Secure distributed systems rely on \emph{trust}. A security assumption
defines the failures and attacks that can be tolerated and names conditions
under which the system may operate.  Implicitly, this determines the trust
in certain components to be correct. In fault-tolerant replicated systems,
trust has traditionally been expressed globally, through a \emph{symmetric}
assumption on the number or kind of faulty processes, which is shared by
all processes.  An example of this is the well-known threshold fault
assumption: the system tolerates up to a finite and limited number of
faulty processes in the system; no guarantees can be given beyond this
about the correct execution of protocols.  More generally, a symmetric
trust assumption is defined through a \emph{fail-prone system}, which is a
collection of subsets of processes, such that each of them contains all the
processes that may at most fail together during a protocol execution.

\emph{Quorum systems}~\cite{DBLP:journals/siamcomp/NaorW98} complement the
notion of fail-prone systems and are used within distributed fault-tolerant
protocols to express trust assumptions operationally.

In the classical interpretation, a quorum system is a collection of subsets
of processes, called \emph{quorums}, with two properties, formally known as
\emph{consistency} and \emph{availability}, respectively, that any two
quorums have a non-empty intersection and that in every execution, there
exists a quorum made of correct processes.  \emph{Byzantine quorum systems}
(BQS) have been formalized by Malkhi and
Reiter~\cite{DBLP:journals/dc/MalkhiR98} and generalize classical quorum
systems by tolerating Byzantine failures, i.e., where faulty processes may
behave arbitrarily.  They are the focus of this work and allow for building
secure, trustworthy systems.  A BQS assumes one global shared Byzantine
fail-prone system and, because of that, use the model of symmetric trust.
Consistency for a BQS demands that any two quorums intersect in a set that
contains at least one correct process in every execution.

Motivated by the requirements of more flexible trust models, particularly
in the context of blockchain networks, new approaches to trust have been
explored.  It is evident that a common trust model cannot be imposed in an
open and decentralized or permissionless environment.  Instead, every
participant in the system should be free to choose who to trust and who not
to trust.  Damg{\aa}rd \emph{et
  al.}~\cite{DBLP:conf/asiacrypt/DamgardDFN07}, and Cachin and
Tackmann~\cite{DBLP:conf/opodis/CachinT19} extend Byzantine quorum systems
to permit subjective trust by introducing \emph{asymmetric} Byzantine
quorum systems.  They let every process specify their own fail-prone system
and quorum system.  Global system guarantees can be derived from these
personal assumptions.  Extending traditional Byzantine quorum systems that
use threshold assumptions, several recent recent
suggestions~\cite{DBLP:conf/osdi/LiuVCQV16,DBLP:conf/icdcn/HowardCM21,DBLP:conf/ccs/MalkhiN019}
have also introduced more flexible notions of trust.

In this paper, we study the problem of composing trust assumptions, as
expressed by symmetric and by asymmetric Byzantine quorum systems.
Starting from two or more running distributed systems, each one with its
own assumption, how can they be combined, so that their participant
groups are joined and operate together?  A simple, but not so intriguing
solution could be to stop all running protocols and to redefine the trust
structure from scratch, with full knowledge of all assumptions across the
participants.  With symmetric trust, a new global assumption that includes
all participants would be defined.  In the asymmetric-trust model, every
process would specify new personal assumptions on all other participants.
Subsequently, the composite system would have to be restarted.  Although
this solution can be effective, it requires that all members of each
initial group express assumptions about the trustworthiness of the
processes in the other groups.  In realistic scenarios, this might not be
possible, since the participants of one system lack knowledge about the
members of other systems, and can therefore not express their trust about
them.  Moreover, one needs to ensure that the combined system satisfies the
liveness and safety conditions, as expressed by the $B^3$-condition for
quorum intersection.  Since the assumptions are personal, it is not
guaranteed, and in practice quite challenging, that the composite system
will indeed satisfy the $B^3$-condition.

This work formulates the problem of composing quorum systems and gives
methods for assembling trust assumptions from different, possibly disjoint,
systems to a common model.  We do so by introducing composition rules for
trust assumptions, in both the symmetric-trust and asymmetric-trust model.
Our methods describe the resulting fail-prone systems and the corresponding
quorum systems.

In a different line of work, subjective trust assumptions have also been
introduced with the Stellar
blockchain~(\url{www.stellar.org})~\cite{Mazieres16,
  DBLP:conf/sosp/LokhavaLMHBGJMM19, DBLP:conf/wdag/LosaGM19}, a
cryptocurrency ranked in the top-20 by market capitalization today.  In
contrast to the original, well-understood notion of quorum systems, these
works depart from the classical intersection requirement among quorums.
Such systems may fork into separate \emph{consensus clusters}, each one
satisfying agreement and liveness on its own.  This implies that consensus
may hold only ``locally'', and a unique consensus across disjoint clusters
is not possible.  More specifically, Losa~\emph{et al.} prove~\cite[Lemma
4]{DBLP:conf/wdag/LosaGM19} that no quorum-based algorithm can guarantee
agreement between two processes whose quorums do not intersect in their
model.  Our work overcomes this impossibility and shows that consensus can
be reached even with disjoint sets of participants, whose trust assumptions
do not intersect.  Moreover, we use the established notion of quorums,
which enables to run many well-understood protocols, such as consensus,
reliable broadcast, emulations of shared memory, and
more~\cite{DBLP:conf/opodis/CachinT19, DBLP:journals/corr/abs-2005-08795}.

Specifically, the contributions of this work are as follows:
\begin{enumerate}
\item We show how to join together two or more systems in a way where
  processes in one system do not need a complete knowledge of the trust
  assumptions of those in the other.
\item We allow processes in each system to maintain their trust
  assumptions within their original system.
\item We define a deterministic rule to extend the trust assumptions of
  each system by including the new participants.
\item Our composition rules guarantee that \emph{consistency} and
  \emph{availability} will be satisfied in the composite quorum system.
\end{enumerate}

\paragraph{Organization.}
The remainder of this work is structured as follows. In Section~\ref{sec:related-works} we review related work. We present our system model and preliminaries on quorum systems with symmetric and asymmetric assumptions in Section~\ref{sec:system-model}.  In Section~\ref{sec:comp_sym} we focus on the symmetric-trust model and show different composition rules on both fail-prone systems and quorum systems. These rules achieve different properties, which we explore formally. A composition rule in the asymmetric model is presented in Section~\ref{sec:comp_asym}. For this proof, we make use of a deterministic method called \emph{purification}, whose purpose is to streamline and improve the trust assumption of each participant in a system, making the composition between more systems possible. We then discuss the implications and the limits of this approach and offer ideas on how to implement our results. Finally, conclusions are drawn in Section~\ref{sec:conclusions}.

\section{Related work}
\label{sec:related-works}

\emph{Byzantine quorum systems} (BQS) have originally been formalized by Malkhi and Reiter~\cite{DBLP:journals/dc/MalkhiR98} to generalize classical quorum systems toward processes prone to Byzantine failures. They model symmetric trust, where every process in the system adheres to a global, common assumption.
Many distributed protocols employ BQS as their foundation; in the area of state-machine replication, for example, they range from PBFT~\cite{DBLP:journals/tocs/CastroL02} to Tendermint~\cite{DBLP:journals/corr/abs-1807-04938}, HotStuff~\cite{DBLP:conf/podc/YinMRGA19}, and other blockchain-specific protocols. Recently, also generalized BQS have been demonstrated for implementing consensus~\cite{DBLP:conf/srds/AlposC20}.

\emph{Measures of quality} for classical (non-Byzantine) quorum system have been studied by Naor and Wool~\cite{DBLP:journals/siamcomp/NaorW98} in terms of the load, capacity, and availability properties. The load (the probability of access of the busiest process) and availability (probability of some quorum surviving failures) properties have then been considered by Malkhi \emph{et al.}~\cite{DBLP:journals/siamcomp/MalkhiRW00} in the context of the Byzantine quorum systems. They construct different types of Byzantine quorum systems with optimal load or availability.

Subsequent literature extends the BQS model, seeking to overcome some limitations and to take them into practice. To this end, \emph{probabilistic} quorum systems have been introduced by Malkhi \emph{et al.}~\cite{DBLP:journals/iandc/MalkhiRWW01} as a tool for ensuring consistency of replicated data with high probability despite both benign and Byzantine failure of processes. They define the \emph{$\epsilon$-intersecting quorum systems} by relaxing the intersection property of a quorum system in a way that every two quorums fail to intersect with some small probability $\epsilon$. By the quality measures, these new quorums show an improvement over the classic and Byzantine ones. 

Alvisi \emph{et al.}~\cite{DBLP:conf/dsn/AlvisiPMRW00} introduce \emph{dynamic} Byzantine quorum systems in the context of quorum-based Byzantine fault-tolerant data services. They present protocols for dynamically changing the threshold of the system. In this this way, they solve an intrinsic limitation of standard Byzantine quorums, which is their dependence on \emph{a-priori} defined resilience thresholds. 

Malkhi \emph{et al.}~\cite{DBLP:conf/ccs/MalkhiN019} define \emph{flexible Byzantine quorums} that allow processes in the system to have different faults models. This work presents a new approach for designing Byzantine fault-tolerant consensus protocols which guarantees higher resilience by introducing a new \emph{alive-but-corrupt} fault type, which denotes processes that attack safety but not liveness.

Recent work has explored frameworks that loosen the global model of trust, allowing processes to choose in a subjective way who to trust.
Damg{\aa}rd \emph{et al.}~\cite{DBLP:conf/asiacrypt/DamgardDFN07} define the basics of \emph{asymmetric trust} for secure computation protocols. Under this model, processes are free to make their personal assumptions regarding other processes, resulting in a broader and richer trust structure, compared to the symmetric model. They introduce a wider class of correct processes, differentiated according to their trust choices. Moreover, they show protocols for synchronous broadcast, verifiable secret sharing, and other primitives. Properties of these protocol can be guaranteed only to a specific subset of correct processes.

 \emph{Asymmetric Byzantine quorum systems} have been introduced by Cachin and Tackmann~\cite{DBLP:conf/opodis/CachinT19} as a natural extension of symmetric Byzantine quorum systems~\cite{DBLP:journals/dc/MalkhiR98} to the model with asymmetric trust. They present protocols for asynchronous Byzantine consistent broadcast, reliable broadcast, and emulations of shared memory with asymmetric quorums. Their work gives rise to a new structure called a \emph{guild}, which is a subset of processes that are called \emph{wise} because they correctly anticipated the actual faults. Some protocol guarantees can only be ensured for wise processes or only for those in a guild. Asymmetric Byzantine consensus protocols have been described as well~\cite{DBLP:journals/corr/abs-2005-08795}.

With the rise of blockchains, protocols using flexible trust structures have been deployed in practice as well. Ripple~(\url{www.ripple.com}) and Stellar~(\url{www.stellar.org}) do not base their resilience guarantees on a global threshold, but allow participants to express their own beliefs. However, their formalization is not a generalization of the BQS model. In this work, we explore compositions of symmetric and asymmetric BQS that are based on the well-studied notions.

A related form of recursive composition of (Byzantine) quorum systems has been explored and utilized in the literature. The idea is that, given two systems, each occurrence of a process in the first is \emph{replaced} by a copy of the second system. Malkhi \emph{et al.}~\cite{DBLP:journals/siamcomp/MalkhiRW00} construct and study composite BQS, such as \emph{recursive threshold} BQS, using this idea. Hirt and Maurer~\cite{DBLP:journals/joc/HirtM00} use this technique to reason about multiparty computation over access structures. Our approach is orthogonal to these works, in the sense that it places the two original systems on the same level. In other words, we explore the failures that two systems can tolerate when they are joined together, as opposed when one is inserted into the other.

\section{System model and preliminaries}
\label{sec:system-model}

\subsection{System model}

\paragraph*{Processes.}
We consider a system $\CP$ with an arbitrary number of \emph{processes} $p_i$, also called \emph{participants}, that communicate with each other.  A protocol for \CP consists of a collection of programs with instructions for all processes. 

\paragraph*{Executions and faults.}
An \emph{execution} starts with all processes in a special initial state; subsequently the processes repeatedly change their state through computation steps. Every execution is fair in the sense that, informally, processes do not halt prematurely when there are still steps to be taken.

A process that follows its protocol during an execution is called \emph{correct}.  On the other hand, a \emph{faulty} process may crash or deviate arbitrarily from its specification, e.g., when \emph{corrupted} by an adversary; such processes are also called \emph{Byzantine}.  We consider only Byzantine faults here and assume for simplicity that the faulty processes fail right at the start of an execution.

\subsection{Preliminaries}
\label{ssec:preliminaries}

We start by presenting definitions and main results in the symmetric-trust model. These will be used in the next section to construct and prove our composition rules.

\begin{definition}[Fail-prone system]\label{def:failprone}
Let \CP be a set of processes. A \emph{fail-prone system} $\CF$ is a collection of subsets of \CP, none of which is contained in another, such that some $F \in \CF$ with $F \subseteq \CP$ is called a \emph{fail-prone set} and contains all processes that may at most fail together in some execution. 

\end{definition}

A complementary structure to the fail-prone system is given by a Byzantine quorum system~\cite{DBLP:journals/dc/MalkhiR98}, defined as follows.

\begin{definition}[Byzantine quorum system]\label{def:quorum}
Let \CP be a set of processes and let $\CF \subseteq 2^{\CP}$ be a \emph{fail-prone system}.
  A \emph{Byzantine quorum system} (BQS) for \CF is a collection of sets of
  processes $\CQ \subseteq 2^{\CP}$, where each $Q \in \CQ$ is called a
  \emph{quorum}, such that:
  \begin{description}
  \item[Consistency:] \[\forall Q_1, Q_2 \in \CQ , \forall F \in \CF: \, Q_1 \cap Q_2 \not
    \subseteq F.\]
\item[Availability:] \[\forall F \in \CF: \, \exists~Q \in \CQ: \, F \cap Q = \emptyset.\]
  \end{description}
\end{definition}

A link between the above two definition is given by the following results.

\begin{definition}[$Q^3$-condition]\label{def:q3}
  Let \CF be a fail-prone system. We say that \CF satisfies the \emph{$Q^3$-condition}, abbreviated
  as $Q^3(\CF)$, if it holds
  \[
    \forall F_1, F_2, F_3 \in \CF: \, \CP \not\subseteq F_1 \cup F_2 \cup F_3.
  \]
\end{definition}

\begin{lemma}[Symmetric quorum system existence~\cite{DBLP:journals/dc/MalkhiR98}]\label{lem:canon}
  Let \CF be a fail-prone system. A Byzantine quorum system for \CF exists
  if and only if~$Q^3(\CF)$. In particular, if $Q^3(\CF)$ holds, then
  $\overline{\CF}$, the bijective complement of \CF, is a Byzantine quorum
  system called \emph{canonical quorum system} of $\CF$.
\end{lemma}

Finally, we present the asymmetric-trust model as introduced by Damg{\aa}rd \emph{et al.}~\cite{DBLP:conf/asiacrypt/DamgardDFN07} and Cachin and Tackmann~\cite{DBLP:conf/opodis/CachinT19}.

\begin{definition}[Asymmetric fail-prone system]\label{def:asymfailprone}
An asymmetric fail-prone system $\BF = [\CF_1, \dots, \CF_n]$ consists of an array of fail-prone systems, where $\CF_i \subseteq 2^{\CP}$ denotes the trust assumption of $p_i$. We assume $p_i \not\in \CF_i$
\end{definition}

One often assumes that $\forall F \in \mathcal{F}_i: p_i \notin F$ for practical reasons, but this is not necessary.
For a system $\CA \subseteq 2^\CP$, let $\CA^*= \{ A' | A' \subseteq A, A \in \CA \}$ denote the collection of
all subsets of the sets in $\CA$.

\begin{definition}[Asymmetric Byzantine quorum system]\label{def:asymquorum}
  Let $\BF = [\CF_1, \dots, \CF_n]$ be an asymmetric fail-prone system. An
  \emph{asymmetric Byzantine quorum system} (ABQS) for \BF is an array of
  collections of sets $\BQ = [\CQ_1, \dots, \CQ_n]$, where
  $\CQ_i \subseteq 2^{\CP}$ for $i \in [1,n]$.  The set
  $\CQ_i \subseteq 2^{\CP}$ is called the \emph{quorum system of $p_i$} and
  any set $Q_i \in \CQ_i$ is called a \emph{quorum (set) for $p_i$}
  whenever the following conditions hold:
  \begin{description}
  \item[Consistency:]
    $\forall i,j \in [1,n]$
\begin{equation*}
 \forall Q_i \in \CQ_i, \forall Q_j \in \CQ_j,
      \forall F_{ij} \in {\CF_i}^* \cap {\CF_j}^*: 
 Q_i \cap Q_j \not\subseteq F_{ij}. 
\end{equation*}    
  \item[Availability:] 
    $\forall i \in [1,n]$
    \[
      \forall F_i \in \CF_i: \, \exists~Q_i \in \CQ_i: \, F_i \cap Q_i =
      \emptyset.
    \]
  \end{description}
\end{definition}

The following property generalizes the $Q^3$-condition from
Definition~\ref{def:q3} to the asymmetric-trust model.

\begin{definition}[$B^3$-condition~\cite{DBLP:conf/asiacrypt/DamgardDFN07,DBLP:conf/opodis/CachinT19}]
\label{def:b3}
  Let \BF be an asymmetric fail-prone system. We say that \BF satisfies the
  \emph{$B^3$-condition}, abbreviated as $B^3(\BF)$, whenever it holds for
  all $i,j \in [1,n]$ that
  \[
    \forall F_i \in \CF_i, \forall F_j\in\CF_j,
    \forall F_{ij} \in {\CF_i}^*\cap{\CF_j}^*: \,
    \CP \not\subseteq F_i \cup F_j \cup F_{ij}. 
  \]
\end{definition}

\begin{lemma}[Asymmetric quorum system existence~\cite{DBLP:conf/opodis/CachinT19}]\label{lem:canon-asym}
An asymmetric fail-prone system $\BF$ satisfies $B^3(\BF)$ if and only if there exists an asymmetric quorum system for $\BF$.
\end{lemma}

For a given asymmetric fail-prone system, we call the list of canonical quorum systems of all processes an \emph{asymmetric canonical quorum system}.

Given a protocol execution with asymmetric Byzantine quorum systems, where $F$ is the actual failed set, the processes are classified in three different types:
\begin{description}
\item[Faulty:] A process $p_i \in F$ is \emph{faulty}.
\item[\Naive:] A correct process $p_i$ for which $F \not\in {\CF_i}^*$
  is called \emph{\naive}.
\item[Wise:] A correct process $p_i$ for which $F \in {\CF_i}^*$ is called
  \emph{wise}.
\end{description}

Recall that all processes are wise under a symmetric trust assumption. Protocols for asymmetric quorums cannot guarantee the same properties for \naive processes as for wise ones.

A useful notion for ensuring liveness and consistency for protocols is that of a \emph{guild}. This is a set of wise processes that contains at least one quorum for each member.

\begin{definition}[Guild]
  Given a fail-prone system \BF, an asymmetric quorum system \BQ for \BF,
  and a protocol execution with faulty processes~$F$, a \emph{guild \CG for
    $F$} satisfies two properties:
  \begin{description}
  \item[Wisdom:] \CG consists of wise processes, i.e.,
    \[
      \forall p_i \in \CG :\, F \in {\CF_i}^*.
    \]
  \item[Closure:] \CG contains a quorum for each of its members, i.e.,
    \[
      \forall p_i \in \CG, \exists~Q_i \in \CQ_i :\, Q_i \subseteq \CG.
    \]
  \end{description}
\end{definition}

Observe that the union of two guild is again a guild~\cite{DBLP:journals/corr/abs-2005-08795}. Every execution with a guild contains a unique \emph{maximal guild}. 

\begin{lemma}[~\cite{DBLP:journals/corr/abs-2005-08795}]\label{lem:quorumnaive}
  Let $\CG$ be the guild for a given execution and let
  $p_i$ be any correct process. Then, every quorum for $p_i$ contains at
  least one process from the guild.
\end{lemma}

\section{Composition of symmetric BQS}\label{sec:comp_sym}
Given two Byzantine quorum systems $\CQ_1$ defined on processes $\CP_1$ with fail-prone system $\CF_1$,
and $\CQ_2$ defined on processes $\CP_2$ with fail-prone system $\CF_2$, we want to provide a \emph{composition} rule between the two that allows the resulting BQS $\CQ_3$ defined on processes $\CP_3 = \CP_1 \cup \CP_2$ with fail-prone system $\CF_3$ to run a distributed protocol together. The resulting system should satisfy the consistency and availability properties of a BQS,
that is, it should remain consistent and live in any execution where a fail-prone set in $\CF_3$ fails. 

In this work we explore the composition of two BQS as a means to allow them jointly run a protocol, \emph{without} requiring the processes in one BQS to make new trust assumptions about the processes in the other.
This is useful in practice because remodeling trust from scratch would be a manual and uncertain process.
We do not consider the composition as a way to increase their resilience. For example, joining four singleton BQS will result in a system with four processes, none of which is expected to fail. This makes sense if one starts from the trust assumptions of singleton BQS; by definition, the single process it contains never fails. There could be other ways to compose the BQS, but they would require changing the assumptions of each individual BQS and it is subject of future work.

According to the previous discussion, we now state properties that we expect any form of composition for BQS should satisfy. 
Characterizing the failures the composite BQS can tolerate now becomes the challenge because multiple definitions of $\CF_3$ are plausible.  We want
to ensure the following \emph{properties}:
\begin{enumerate}
\item Any $B \in \CF_3$ satisfies $B|_{\CP_1} \in \CF_1^*$, i.e., the failure of $B$ is tolerated in the first system.
\item Any $B \in \CF_3$ satisfies $B|_{\CP_2} \in \CF_2^*$, i.e., the failure of $B$ is tolerated in the second system.
\item $\CF_3$ satisfies the $Q^3$-condition.
\item For any $B \in \CF_3$, there exists a $Q \in \CQ_3$, a quorum system in the composite system, such that $B \cap Q = \emptyset$, i.e., there is always a quorum consisting only of correct processes.
\end{enumerate}
In the text above, the notation $\CX|_\CP$ denotes the restriction of a set \CX to \CP.

We need properties~1 and~2 because, as we shall see next, they imply Property~3, and, hence, ensure consistency for the composite BQS against any fail-prone set in $\CF_3$.
Moreover, they enable a composition by using the existing assumptions, without requiring
a redesign of the two systems.
One might also desire that the inverse of properties~1 and~2 be satisfied, i.e., that any
fail-prone set in $\CF_1$ and $\CF_2$ be tolerated in $\CF_3$. However, we will later see that
this does not always result in a BQS (i.e., in a fail-prone system that satisfies the $Q^3$-condition).
Thus, the objective of a composition rule is to satisfy these properties, thus ensuring safety, while producing a maximal fail-prone system $\CF_3$
(in the sense that it contains the largest fail-prone sets that could be created without having to redefine the trust assumptions within the original systems). 
Finally, the composition rule should also satisfy Property~4, which ensures liveness in the composite system.

\begin{lemma}\label{lem:properties_q3}
Properties~1 and~2 above imply Property~3.
\end{lemma}
\begin{proof}
Let us assume that $Q^3(\CF_1)$ and $Q^3(\CF_2)$. Towards a contradiction, let $F_A, F_B, F_C \in \CF_3$ such that $F_A \cup F_B \cup F_C = \CP_3$.
Now consider the restriction of $F_A, F_B$ and $F_C$ to $\CP_1$ (and similarly to $\CP_2$).
We have that $F_A|_{\CP_1} \cup F_B|_{\CP_1} \cup F_C|_{\CP_1} = \CP_1$.
However, from Property~1, the sets $F_A|_{\CP_1}$, $F_B|_{\CP_1}$, and $F_C|_{\CP_1}$ are each (subsets of) fail-prone sets in $\CF_1$.
We thus have found three fail-prone sets that cover $\CP_1$, a contradiction to $\CF_1$ satisfying the $Q^3$-condition.
\end{proof}

With this list of goals, we now proceed to specific constructions.
In the following, we present three composition methods of increasing suitability and give examples to show their weaknesses and strengths.

\begin{construction}[Union composition]\label{def:comp_sym_2}
Let $\CQ_1$ be a BQS defined on processes $\CP_1$ with fail-prone system $\CF_1$, and $\CQ_2$ a BQS defined on processes $\CP_2$ with fail-prone system $\CF_2$,
where $\CP_1 \cap \CP_2 = \emptyset$.
The \emph{union composition} of $\CQ_1$ and $\CQ_2$ is a system defined on processes $\CP_3 = \CP_1 \cup \CP_2$
with fail-prone system
\begin{equation*}
\CF_3 = \CF_1 \cup \CF_2.
\end{equation*}
\end{construction}

We can easily verify that the previous definition, given that $\CP_1 \cap \CP_2 = \emptyset$, fulfills Properties~1 and~2.
Thus, $\CF_3$ satisfies the $Q^3$-condition and there exists a BQS $\CQ_3$ with fail-prone system $\CF_3$. 

\begin{lemma}
\label{lem:proof-q3-sym}
Given $\CF_3$ as in Construction~\ref{def:comp_sym_2}, a BQS $\CQ_3$ is 
$$\CQ_3 = \{ Q_i \cup Q_j \mid Q_i \in \CQ_1, Q_j \in \CQ_2 \},$$ with $\CQ_1$ and $\CQ_2$ BQS.
\end{lemma}

\begin{proof}
We first show that consistency property holds. 
So, for every $Q_1, Q_2 \in \CQ_3$ such that $Q_1 = Q_i \cup Q_j$ and $Q_2=Q^{'}_i \cup Q^{'}_j$,
with $Q_i,Q^{'}_i \in \CQ_1$ and $Q_j, Q^{'}_j \in \CQ_2$,
and for every $F \in \CF_3$, with $F \in \CF_1$ or $F \in \CF_2$, we have $Q_1 \cap Q_2 = (Q_i \cup Q_j) \cap (Q^{'}_i \cup Q^{'}_j) $, which equals
$(Q_i  \cap Q^{'}_i ) \cup (Q_j \cap Q^{'}_j )$, because $\CP_1 \cap \CP_2  = \emptyset$.
By assumption, both $\CQ_1$ and $\CQ_2$ are BQS. 
This means that, if $F \in \CF_1$, then $\CQ_i \cap Q^{'}_i \not\subseteq F$, 
and if $F \in \CF_2$, then $\CQ_j \cap Q^{'}_j \not\subseteq F$.
The property then follows.
Finally, the availability property follows from the fact that $\CP_1$ and $\CP_2$ are disjoint and $\CQ_1$ and $\CQ_2$ are BQS.
\end{proof}

However, the fail-prone system obtained by Construction~\ref{def:comp_sym_2} results in a fail-prone system that tolerates only a few failures, namely those tolerated in each of the two original systems, and not any combination of them. Moreover, it would not work if $\CP_1$ and $\CP_2$ had any processes in common.
The next notion moves towards a composition that tolerates any combination of failures that would be tolerated in the original systems. 

\begin{construction}[Cartesian composition on disjoint sets]\label{def:comp_sym_1}
Let $\CQ_1$ be a BQS defined on processes $\CP_1$ with fail-prone system $\CF_1$, and $\CQ_2$ a BQS defined on processes $\CP_2$ with fail-prone system $\CF_2$,
where $\CP_1 \cap \CP_2 = \emptyset$.
Then the \emph{Cartesian composition} of $\CQ_1$ and $\CQ_2$ is defined on processes $\CP_3 = \CP_1 \cup \CP_2$
and tolerates the failure of any combination of fail-prone sets of the original BQS. Formally,
\[
\CF_3 = \{ F_i \cup F_j \mid F_i \in \CF_1, F_j \in \CF_2 \}.
\]
\end{construction}

\begin{lemma}
\label{lem:cart-prod}
If $Q^3(\CF_1)$ and $Q^3(\CF_2)$, then for the fail-prone system $\CF_3$ according to Construction~\ref{def:comp_sym_1}, $Q^3(\CF_3)$.
\end{lemma}
\begin{proof}
Any $B \in \CF_3$ satisfies $B|_{\CP_1} \in \CF_1$ and $B|_{\CP_2} \in \CF_2$, since $\CP_1 \cap \CP_2 = \emptyset$.
Hence, the composition in Definition~\ref{def:comp_sym_1} satisfies the Properties~1 and~2, and, by Lemma~\ref{lem:properties_q3}, also $Q^3(\CF_3)$.
\end{proof}

The previous lemma implies the existence of a BQS $\CQ_3$ with fail-prone system $\CF_3$. 
Such a $\CQ_3$ can be obtained, as earlier, by
$$\CQ_3 = \{ Q_i \cup Q_j \mid Q_i \in \CQ_1, Q_j \in \CQ_2 \}.$$

It is easy to show, in a similar way as in Lemma~\ref{lem:proof-q3-sym}, that this $\CQ_3$ satisfies consistency and availability properties. 
Moreover, if $\CQ_1$ and $\CQ_2$ are canonical, $\CQ_3$ will be the canonical BQS for $\CF_3$.

\begin{example}\label{ex:threshold}
Let us consider the threshold case. Suppose $\CQ_1$ and $\CQ_2$ be two BQS, defined on $\CP_1$ and $\CP_2$, where $\CP_1 \cap \CP_2 = \emptyset$, containing $7$ and $10$ processes, and tolerating the failure of any $2$ and $3$ processes, respectively. This means that the first fail-prone system contains ${7 \choose 2}=21$ sets of processes and the second fail-prone system contains ${10 \choose 3}=120$ sets. 
Because in this work we join systems with already existing failure assumptions, we refrain from changing these assumptions for the composite system. Nevertheless, according to Lemma~\ref{lem:cart-prod}, the Cartesian product of the fail-prone systems leads to a fail-prone system where the $Q^3$-condition holds, assuming that the starting systems both satisfy the $Q^3$-condition and are disjoint.

We apply Construction~\ref{def:comp_sym_1} here, observing that the $Q^3$-condition is the generalization of the condition $n>3f$ for the threshold case.
As a result we obtain an assumption on $17$ processes, which tolerates the failure of $5$ processes, where $2$ processes are from $\CP_1$ and $3$ from $\CP_2$. 
More formally, the failure of a set $F$ is tolerated in the composite system if and only if
$\left| F \cap {P_1} \right| \leq 2 \land \left| F \cap {P_2} \right| \leq 3$.

This gives a total of $2520$ possible tolerated subsets. Observe that $\CQ_3$ is not a threshold BQS any more, and this was intended. A threshold BQS made of $17$ processes would tolerate the failure of any $5$ processes, which would lead to a total of ${17 \choose 5}=6188$ fail-prone sets. 
\end{example}

\begin{example}\label{ex:cartes-inters}
We now show how Construction~\ref{def:comp_sym_1} fails to create a BQS $\CQ_3$ if $\CP_1$ and $\CP_2$ intersect,
because the $Q^3$-condition may not hold in the composite system.
Let $\CQ_1$ defined on $\CP_1 = \{ a, b, c, d, e \}$ with fail-prone system $\CF_1 = \{ \{a\}, \{b, c\}, \{d\}, \{c, e\} \}$
and $\CQ_2$ defined on $\CP_2 = \{ d, e, f, g, h \}$ with fail-prone system $\CF_2 = \{ \{d\}, \{e\}, \{f,g\}, \{h\} \}$.

It is easy to verify that the $Q^3$-condition is satisfied in $\CQ_1$ and $\CQ_2$.
We also see that, according to Construction~\ref{def:comp_sym_1}, $\CQ_3$ with processes $\CP_3=\CP_1 \cup \CP_2$
contains, among others, the fail-prone sets $\{ a, f, g \}, \{ b, c, h \}, \{c, e, d \}$, which cover $\CP_3$.
Consequently, $\CQ_3$ is not a BQS.
\end{example}

Example~\ref{ex:cartes-inters} shows that the Cartesian composition among fail-prone systems does not lead to a fail-prone system where the $Q^3$-condition holds, if the two systems have common processes. To overcome this issue, we introduce a third construction.

\begin{definition}\label{def:cart_operator}
Let $\CA=\{A_1,\ldots,A_m\}$ and $\CB=\{B_1,\ldots,B_n\}$ be two sets of subsets of $\CP_1$ and $\CP_2$, respectively.
We define $\CA \otimes \CB$ as the set
that contains the union of all sets $A_i \in \CA^{*}$ and $B _j\in \CB^{*}$, under the restriction that either both $A_i$ and $B_j$ contain exactly the same subset of the processes common to $\CP_1$ and $\CP_2$ or they do not have anything in common.
Formally,
\begin{multline*}
\CA \otimes \CB =
 \bigl\{A_i \cup B_j \mid A_i \in \CA^{*} \land B_j \in \CB^{*} \land
  (\forall C \subseteq \CP_1 \cap \CP_2 : C \subseteq A_i \Leftrightarrow C \subseteq B_j) \bigr\}.  
\end{multline*}
\end{definition}

\begin{construction}[Cartesian composition]\label{def:comp_sym_correct}
Let $\CQ_1$ be a BQS defined on processes $\CP_1$ with fail-prone system $\CF_1$ and $\CQ_2$ a BQS defined on processes $\CP_2$ with fail-prone system $\CF_2$,
where $\CP_1$ and $\CP_2$ might contain common processes.
Then the \emph{composition} of $\CQ_1$ and $\CQ_2$ is defined on $\CP_3 = \CP_1 \cup \CP_2$
and tolerates the failure of any combination of any fail-prone set (or subset of it) of the first system and any fail-prone set (or subset) of the second system, such that both contain exactly the same subset of the common processes. Formally,
\begin{multline*}
\CF_3 = \CF_1 \otimes \CF_2 =
  \bigl\{ F_i \cup F_j  ~|~ F_i \in \CF_1^* \land F_j \in \CF_2^* \land
(\forall C \subseteq \CP_1 \cap \CP_2 : C \subseteq F_i \Leftrightarrow C \subseteq F_j) \bigr\}.  
\end{multline*}

\end{construction}
The rule of Construction~\ref{def:comp_sym_correct} states that any fail-prone set in $\CF_3$ is of the form $F_i \cup F_j$, where $F_i$ and $F_j$ are fail-prone sets (or subsets of fail-prone sets) that either do not have any processes in common or, if they do, both contain exactly the same subset of $\CP_1 \cup \CP_2$. We demand $F_i \in \CF_1^*$  and $F_j \in \CF_2^*$, instead of $F_i \in \CF_1$ and $F_j \in \CF_2$, in order to construct a maximal $\CF_3$, in the sense that it contains the maximal fail-prone sets that satisfy Properties~1 and~2.

\begin{lemma}\label{lem:comp_sym_correct}
If $Q^3(\CF_1)$ and $Q^3(\CF_2)$, then $Q^3(\CF_3)$, with $\CF_3$ as in Construction~\ref{def:comp_sym_correct}.
\end{lemma}
\begin{proof}
Any $B \in \CF_3$ either does not contain a set of common processes $C$ among $\CP_1$ and $\CP_2$ or it does. 
In the former case, it is immediate to see that $B|_{\CP_1} \in \CF_1^*$ and $B|_{\CP_2} \in \CF_2^*$. 
In the latter case, $B$ has been created as the union between $F_i \in \CF_1^*$ and $F_j \in \CF_2^*$, both containing the same subset of $P_1 \cap P_2$, according to Construction~\ref{def:comp_sym_correct}. 
It is thus not possible that a new element of $\CP_1$ appears in $B|_{\CP_1}$ that was not already in $F_i$,
and similarly that a new element of $\CP_2$ appears in $B|_{\CP_2}$ that was not already in $F_j$.
This implies that $B|_{\CP_1} \in \CF_1^*$ and $B|_{\CP_2} \in \CF_2^*$, and from Lemma~\ref{lem:properties_q3} we get $Q^3(\CF_3)$.
\end{proof}

\begin{lemma}
\label{lem:q3-sym-2}
Given $\CF_3$ as in Construction~\ref{def:comp_sym_correct}, a BQS $\CQ_3$ is
$$\CQ_3 = \{ Q_i \cup Q_j \mid Q_i \in \CQ_1, Q_j \in \CQ_2 \},$$
with $\CQ_1$ and $\CQ_2$ BQS.
\end{lemma}
\begin{proof}
Consistency and availability properties of $\CQ_3$ can be proved in a similar way as Lemma~\ref{lem:proof-q3-sym}, assuming $\CQ_1$ and $\CQ_2$ to be BQS. In fact, as in Lemma~\ref{lem:proof-q3-sym}, we have that for every $Q_1, Q_2 \in \CQ_3$, such that $Q_1 = Q_i \cup Q_j$ and $Q_2=Q^{'}_i \cup Q^{'}_j$,
with $Q_i,Q^{'}_i \in \CQ_1$ and $Q_j, Q^{'}_j \in \CQ_2$, we have $Q_1 \cap Q_2 = (Q_i \cup Q_j) \cap (Q^{'}_i \cup Q^{'}_j) $, which results in $(Q_i \cap Q^{'}_i) \cup (Q_i \cap Q^{'}_j) \cup (Q_j \cap Q^{'}_i) \cup (Q_j \cap Q^{'}_j).$ If $\CP_1 \cap \CP_2 = \emptyset$, it is trivial to prove the result. Otherwise, given $F \in \CF_3$ with $F = F_i \cup F_j$, where $F_i \in \CF_1^* \land F_j \in \CF_2^* \land \forall C \subseteq \CP_1 \cap \CP_2 : C \subseteq F_i \Leftrightarrow C \subseteq F_j$, we have two cases. If there are no common processes between $F_i$ and $F_j$, then observe that $F_i$ is contained in $\CF^{*}_i$ and it is then a subset of a fail-prone set $\overline{F}_i$ in $\CF_i$. The same happens for $F_j$. By assumptions, $\CQ_1$ (respectively, $\CQ_2$) are BQS. It follow that, $(Q_i \cap Q^{'}_i)$ (respectively, $(Q_j \cap Q^{'}_j)$) is not a proper subset of $\overline{F}_i$ and consequently of $F_i$ (respectively of $F_j$). The result follows. The same reasoning can be applied if $F_i$ and $F_j$ contain a common subset $C \subseteq \CP_1 \cap \CP_2$.
\end{proof}

\begin{example}
Let us consider again the threshold case, where $\CQ_1$ is defined on participants $\CP_1 = \{a,b,c,d,e,f,g \}$ and $\CQ_2$ on $\CP_2 = \{g,h,i,j,k,l,m,n,o,p \}$.
According to Construction~\ref{def:comp_sym_correct}, any two processes in $\CP_1$ together with any three processes in $\CP_2$ are tolerated to fail,
because these failures would be tolerated in the original systems. However, if $g$ together with any other process in $\CP_1$ fails, then only two more failures
in $\CP_2$ are tolerated, because $g \in \CP_2$ has already failed in the first system.
\end{example}

\begin{example}
\label{ex:can_quor}
Let $\CQ_1$ be defined on processes $\CP_1 = \{ a, b, c, d, e \}$ and with fail-prone system $ \CF_1 = \{ \{a\}, \{b, c\}, \{d\}, \{c, e\} \}$ and $\CQ_2$ be defined on processes $\CP_2 = \{ d, e, f, g, h \}$ with fail-prone system $\CF_2 = \{ \{d\}, \{e\}, \{f, g\}, \{h\} \}$.  Then, according to Construction~\ref{def:comp_sym_correct} processes in $\CP_3 = \{a, b, c, d, e, f, g, h\}$ have fail-prone system \[\CF_3 = \{\{a, f, g\}, \{a, h\}, \{ b, c, f, g\}, \{ b, c, h\}, \{ d\}, \{ c, e\}\}.\] It is easy to verify that $Q^3(\CF_3)$.
\end{example}

\section{Composition of asymmetric BQS}\label{sec:comp_asym}

We now explore the composition of two asymmetric Byzantine quorum systems. Given two ABQS, $\BQ_1$ defined on processes $\CP_1$ with fail-prone system $\BF_1$, and $\BQ_2$ defined on processes $\CP_2$ with fail-prone system $\BF_2$, we want to provide a \emph{composition} rule that allows the processes $\CP_3 = \CP_1 \cup \CP_2$ to form an ABQS $\CQ_3$ with fail-prone system $\BF_3$.

\subsection{Approaches to asymmetric trust}

In the context of blockchains, different models of asymmetric trust have been proposed, united by a shared principle regarding the subjectivity of truth among the participants, but differentiated by fundamental properties that determine their limitations and strengths. 

In this section, we compare our model of asymmetric trust~\cite{DBLP:conf/opodis/CachinT19} with the model of personal Byzantine quorum system (PBQS) introduced by Losa \emph{et al.}~\cite{DBLP:conf/wdag/LosaGM19}.  PBQS extend and improve the formalization used by Stellar~\cite{Mazieres16, DBLP:conf/sosp/LokhavaLMHBGJMM19} and give a new interpretation of a quorum system for subjective trust.

In a PBQS, each participant has its own notion of a quorum, with the requirement that if $Q_i \subseteq \CP$ is a quorum for a process $p_i$ and $p_j \in Q_i$, then it exists $Q_j$ for $p_j$ such that $Q_j \subseteq Q_i$. In other words, a quorum $Q_i$ for $p_i$ should contain at least one quorum for each $p_j \in Q_i$. Given this definition, a PBQS consists of a set of participants $\CP$, a set of faulty processes $F \subseteq \CP$, a set of correct processes $\CW = \CP \setminus F$, and a function mapping a participant $p_i$ to its non-empty set of quorums. In other words, Losa \emph{et al.} construct a PBQS starting from an arbitrary set $F$. This notion of a quorum system differs also from the well-known formalization~\cite{DBLP:journals/dc/MalkhiR98} because a quorum in a PBQS is a private notion. In the traditional model, all quorums are public and known to every participant. From this, it follows that a global intersection property is absent from PBQS. 

An asymmetric Byzantine quorum system (ABQS, cf. Section~\ref{ssec:preliminaries}) is defined from an asymmetric fail-prone system, which contains the fail-prone systems of every participant, and requires a global intersection property for consistency. ABQS extend the traditional notion of Byzantine quorum systems~\cite{DBLP:journals/dc/MalkhiR98}. With an ABQS, the correct processes can be grouped into \naive and wise ones, depending on their trust assumptions. According to this distinction, one can guarantee most properties of a protocol only to wise processes. 

An useful structure in an ABQS \BQ is a \emph{kernel}~\cite{DBLP:conf/opodis/CachinT19, DBLP:journals/corr/abs-2005-08795} of each quorum system~$\CQ_i$  for $p_i$. This is a set $K_i \subseteq \CP$ with the property that for every $Q \in \CQ_i: \, K \cap Q \neq \emptyset$. In other words, a kernel is a set of processes that intersects every quorum in a quorum system $\CQ_i$ for a process $p_i$; it generalizes sets of size $f + 1$ in the traditional symmetric threshold model. Losa \emph{et al.} define a similar structure called a \emph{blocking} set. In particular, given $\CR$ a set of participants, a process $p_i$ is \emph{blocked} by $\CR$ when every quorum of $p_i$ intersects $\CR$. Moreover, they show that if a process $p_i$ is blocked by the set of faulty processes $F$, then it is impossible to guarantee liveness for $p_i$. With ABQS, this cannot happen: by the availability property of an ABQS, for every set of faulty processes, it always exists a quorum $Q_i \in \CQ_i$ for $p_i$ that consists only of correct processes. It follows that, eventually, a process~$p_i$ will hear from a quorum of processes for itself, even if all the malicious processes remain silent. 

Finally, an ABQS execution gives rise to a \emph{guild}, a set of wise processes that contains at least one quorum for each of its members. The existence of a guild is essential for protocols with ABQS and it also plays a fundamental role for the composition of ABQS. Guilds cannot be disjoint and the union of two guilds is again a guild~\cite{DBLP:journals/corr/abs-2005-08795}. The analogue of a guild in a PBQS is a \emph{consensus cluster}, which is a subset $\CS \subseteq \CW$ such that for every two quorums $Q_i$ and $Q_j$ of some members of $\CS$, it holds $Q_i \cap Q_j \cap \CW \neq \emptyset$, and for every $p_i \in \CS$, it exists a quorum $Q_i$ for $p_i$ such that $Q_i \subseteq \CS$. However, despite these similarities to a guild, two consensus clusters can be disjoint (due to the missing intersection requirement in a PBQS). This implies that for consensus with PBQS, agreement may hold only locally, and achieving  consensus across disjoint clusters may not be possible.

\subsection{The tolerated system of an ABQS}

For defining composition with ABQS, we first introduce the central notion of the \emph{tolerated system} of an ABQS.
Recall that symmetric BQS start from a common understanding of the world; the participants agree on the possible failures, that is, on which participants might crash or collaborate to break security. In an asymmetric BQS, no such common understanding exists, either because there is not enough knowledge to make such an assumption on the system, or because the participants simply do not agree with each other. In this model, every participant expresses its own beliefs and expectations, and no global notion of ``correct'' belief exists. In every execution, however, there will be a ground truth, manifested by a set of actually faulty participants, and not all members of the system will have correctly anticipated this ground truth. Again, since there is no global understanding of the world, this is expected to happen. 
However, the participants might still be able to make progress (where progress is defined by the protocol they are running), exactly in those executions when a guild exists. Recent works on consensus with ABQS have conditioned safety and liveness properties on the existence of such a set. In the context of Byzantine consensus, Cachin and Zanolini~\cite{DBLP:journals/corr/abs-2005-08795} show that a guild is required to solve asynchronous consensus and that consensus properties are guaranteed in all executions with a guild. 

An external party examining an ABQS without any prior knowledge or beliefs about the participants cannot assess the trust assumptions of any individual participant. However, the third party can evaluate the ABQS based on its ability to make progress through a guild.

The central concept for composing two ABQS is the \emph{tolerated system} of an ABQS.  Recall that in an execution where all processes in $B \subset \CP$ actually fail, there may also be \naive processes, wise processes that form a guild $\CG$, and wise processes outside the guild (\cite[Example~1]{DBLP:journals/corr/abs-2005-08795}).  For a specific guild $\CG \neq \emptyset$, the union of all those processes outside \CG is called a \emph{tolerated set} because the guild is autonomous without any of them. Hence, the tolerated set consists of the faulty, the \naive, and the wise processes outside the guild. The tolerated system contains all the tolerated sets. Formally, we have the following definition.

\begin{definition}[Tolerated system]\label{def:tolerated_sys}
  The \emph{tolerated system} $\CT$ of an ABQS  \BQ defined on processes $\CP$ is
\begin{equation*}
 \CT =\{\CP \setminus G \text{, for any possible guild } G \text{ of } \BQ \}.
\end{equation*}  
\end{definition}

Intuitively, the tolerated system of an ABQS reflects the resilience of the ABQS: even without the processes in a tolerated set, there still exists a guild. Therefore, the tolerated system characterizes the executions in which some of the participants in the asymmetric system will be able to operate correctly and make progress. In that sense, the tolerated system of an ABQS is the counterpart of the fail-prone system for a BQS. 

Notice that the tolerated system is a global notion emerging from the subjective trust choices of the participating processes; any party that knows the fail-prone and quorum systems of all processes can calculate it. We show later that the tolerated systems of two ABQS play a crucial role for composing them; the processes in the first system will use the tolerated sets of the second system as their trust assumptions, and vice versa. Consequently, the processes in the first system only need to know the tolerated system of the second system.

The following lemma shows that the tolerated system of a canonical ABQS naturally corresponds to a BQS.

\begin{lemma}\label{lem:b3-q3}
  Let $\BQ$ be an ABQS on processes \CP with asymmetric fail-prone system
  $\BF = \overline{\BQ}$, i.e., such that \BQ is a canonical ABQS.  Then
  the tolerated system \CT of \BQ is a BQS. In particular, if $B^3(\BF)$,
  then $Q^3(\CT)$.
\end{lemma}
\begin{proof}
  Towards a contradiction, let us assume that $\CT$ does not satisfy the $Q^3$-condition. This means that there exist $T_1, T_2, T_3 \in \CT$ such that $T_1 \cup T_2 \cup T_3 = \CP$. Also, let $\CG_1, \CG_2, \CG_3$ be the corresponding guilds, i.e., $\CG_1 = \CP \setminus \CT_1, \CG_2 = \CP \setminus \CT_2$ and $\CG_3 = \CP \setminus \CT_3$. 
Without loss of generality every guild contains at least a process, and at least a quorum for this process is fully contained in the guild. By the consistency property of an ABQS, these quorums must intersect pairwise, hence the guilds also intersect pairwise. This means that there exist processes $p_{i} \in \CG_1 \cap \CG_2$ and $p_{j} \in \CG_2 \cap \CG_3$. Now, because $p_i$ is a member of $\CG_1$, we can make the following reasoning: $p_i$ has a quorum $Q_i \in \CQ_i$ such that $Q_i \subseteq \CG_1$, the BQS is canonical, so $p_i$ has a fail-prone set $F_i = \CP \setminus Q_i \in \CF_i$, thus we get $ T_1 \subseteq F_i$, i.e., $T_1 \in \CF_i$. With similar reasoning, we get $T_2 \in \CF_i$ (because $p_i \in \CG_2$), $T_2 \in \CF_j$ (because $p_j \in \CG_2$), and $T_3 \in \CF_j$ (because $p_j \in \CG_3$). But this is a contradiction, because $p_i$ and $p_j$ with fail-prone sets $T_1, T_2$, and $T_3$ violate the $B^3$-condition in $\BQ$.
\end{proof}

As has been known before, by Lemma~\ref{lem:canon}, if $\CT$ satisfies the $Q^3$-condition, then there exists also a symmetric BQS for the fail-prone system \CT; for instance, this may be the canonical BQS~$\overline{\CT}$. 

Lemma~\ref{lem:b3-q3} confirms the intuition that the tolerated set of an ABQS is the counterpart of a fail-prone set in a BQS.

\subsection{How clients interact with an ABQS}

Many practical replication protocols separate clients from replicas; in state-machine replication, clients submit commands, replicas totally order and execute them, and then send back responses to the clients. When the expected failures among replicas are modeled as a BQS, that is, with a symmetric trust assumption, the clients wait for responses from a quorum of replicas. However, if the trust assumption among the replicas is asymmetric, it is unclear which sets of participants are capable to convince a client to accept a response. The subjective quorums of the replicas only express their personal beliefs, which the clients may not share.

One way to resolve this could be to let each client express trust in the replicas through its own quorum system.  But if clients do not have sufficient knowledge to make such assumptions, they need a global property of the quorum system to decide on its responses, and this can be the tolerated system. Note that every guild formed by replicas corresponds to the complement of a tolerated set. This indicates that (at least some) replicas did agree on their trustworthiness, and this may convince the client. Indeed, we will use this idea in the composition procedure for ABQS. Specifically, the participants of each system may operate as clients of the other and could send a composition-request message, waiting for responses from a guild of participants.

\subsection{Composition of ABQS}

Based on the remarks above, the we claim that any form of composition
between two ABQS must satisfy the following conditions.  Regarding 
notation, we want to compose $\BQ_1$ with $\BQ_2$, resulting in $\BQ_3$,
with respective asymmetric fail-prone systems $\BF_1$, $\BF_2$, and
$\BF_3$.  For $k=1,2$ and for any $p_i \in \CP_k$, let $\CF^{(k)}_i$ be the
fail-prone system of $p_i$ in $\BF_k$, and $\CF^{(3)}_i$ the fail-prone
system of $p_i$ in the resulting $\BF_3$.  Moreover, let $\CT_k$ be the
tolerated system of $\BQ_k$.

\begin{enumerate}
\item If $p_i \in \CP_1$ and $p_i \in \CP_2$, then any
  $F_i \in \CF^{(3)}_i$ must respect the trust assumptions of $p_i$ in
  $\CP_1$ and in $\CP_2$, i.e., it must satisfy
  $F_i|_{\CP_1} \in {\CF^{(1)}_i}^*$ and
  $F_i|_{\CP_2} \in {\CF^{(2)}_i}^*$.  If $p_i$ is only in $\CP_1$ (and the
  same holds for $\CP_2$), then any $F_i \in \CF^{(3)}_i$ must respect the
  assumptions of $p_i$ in $\CP_1$, i.e.,
  $F_i|_{\CP_1} \in {\CF^{(1)}_i}^*$, and $F_i|_{\CP_2}$ can only be one of
  the tolerated sets in $\CP_2$, i.e., $F_i|_{\CP_2} \in {\CT_2}^*$, since
  $p_i$ has no assumptions for $\CP_2$.  This generalizes Properties~1
  and~2 of the symmetric composition.

\item If the $B^3$-condition holds for $\BF_1$ and for $\BF_2$, then it
  also holds for composite system, for $\BF_3$. This is a generalization of
  Property~3 of the symmetric composition.

\item For any $p_i \in \CP_3$ and any $F_i \in \CF^{(3)}_i$, there exists a
  quorum $Q_i \in \CQ^{(3)}_i$, such that $F_i \cap Q_i = \emptyset$.
\end{enumerate}

Up to here, these three properties are generalizations of the corresponding
properties of the symmetric composition. However, in the asymmetric case,
we also want to achieve the following.

\begin{enumerate}\setcounter{enumi}{3}
\item{\textbf{Preserving wisdom.}} In all executions, where there exists a
  guild $\CG_1$ in ABQS $\BQ_1$ and a guild $\CG_2$ in $\BQ_2$, the
  processes in $\CG_1 \cup \CG_2$ will form a guild in $\BQ_3$.  The
  intuition is that, given an execution with $B$ as actual faulty set, if a
  process correctly foresees $B$ (and thus enjoys the properties of a
  guild) in its own system, and if there is a guild in the other system,
  then this process should also enjoy the properties of a guild in the
  composite system.

\item{\textbf{Reducibility to symmetric.}} If all processes have the same
  trust assumptions (in which case $\BQ_1$ and $\BQ_2$ reduce to symmetric
  BQS), then the composite system $\BQ_3$ is a symmetric BQS and satisfies
  the properties of symmetric composition.
\end{enumerate}

\begin{lemma}
\label{lem:p1-p5}
Property~1 implies Property~5.
\end{lemma}
\begin{proof}
  This follows immediately by observing that when all processes in $\CP_k$
  have the same fail-prone system $\CF_k$, for $k = 1, 2$, then the
  tolerated system $\CT_k$ is $\CF_k$ itself. Then, Property~1 implies that
  $\CF^{(3)}_i$ is the same for every $p_i \in \CP_3$, and that every
  $B \in \CF^{(3)}_i$ satisfies $B|_{\CP_1} \in {\CF_1}^*$ and
  $B|_{\CP_2} \in {\CF_2}^*$, which is what Properties~1 and~2 of the
  symmetric composition require.
\end{proof}

Now let us consider two ABQS $\BQ_1$ and $\BQ_2$ on processes $\CP_1$ and
$\CP_2$ with asymmetric fail-prone systems $\BF_1$ and $\BF_2$,
respectively. All processes in $\CP_1$ and $\CP_2$ wish to jointly run a
protocol, without making any extra assumption about the participants of the
other group.  Intuitively speaking, each group might have their own issues,
their own agreements and disagreements, their own good and bad executions,
but they still want to work together. As reasoned earlier, each participant
in $\CP_1$ is an external observer for $\CP_2$. Hence, the best a
participant in $\CP_1$ can do, assuming they have no knowledge, beliefs, or
assumptions for the participants of the second group, is to use the
tolerated system of $\BQ_2$. The same applies, of course, for participants
in $\CP_2$. This leads to the composition procedure we describe next.

\begin{construction}[Purification]
\label{const:pur}
  Let $\BQ$ an ABQS on processes $\CP = \{p_1, \ldots, p_n \}$, with
  asymmetric fail-prone system $\BF = \{\CF_1, \ldots, \CF_n \}$,  such that $B^3(\BF)$,
  and let $\CT$ its tolerated system. Assume $Q^3(\CT)$. As we have seen, this
  is always the case for canonical ABQS.
  We want to \emph{purify} $\BF$ so that
  $B^3([ \CF_1, \ldots, \CF_n, \CT ])$, i.e.,
  $\forall F_i \in \CF_i, \forall F_j \in \CT, \forall F_{ij} \in \CF^{*}_i
  \cap \CT^{*}$ it holds that
  $\CP \not\subseteq F_i \cup F_j \cup F_{ij} $.  To do so, every process
  $p_i$ evaluates the $B^3$-condition including $\CT$ in the
  asymmetric fail-prone system $\BF$.  If it does not hold, then for any
  $F_i \in \CF_i$ that violates the $B^3$-condition, $p_i$ removes $F_i$
  from $\CF_i$, and adds to $\CF_i$ all those subsets of $F_i$ that do not
  violate the $B^3$-condition.  This results in a \emph{purified}
  fail-prone system, which, by construction, satisfies the $B^3$-condition.

  Intuitively, the purification procedure removes fail-prone systems that 
  are ``useless,'' in the sense that they do not influence the existence of a guild,
  as shown by the next lemma.
  Seen from a higher level, it is an expression of the fact that processes have 
  their own beliefs, but also need to adapt to those of the others; a process~$p_i$ might 
  expect a set $F$ to fail during an execution and construct its fail-prone system $\CF_i$ so
  as to be protected against $F$. However, if the beliefs of other
  processes are such that the failure of $F$ does not lead to a guild, i.e., $F$ is not
  tolerated, then $p_i$ can not benefit from including $F$ in $\CF_i$.
\end{construction}

\begin{lemma}
\label{lem:guild-in-pur}
  For every possible execution with a guild \CG, a process in \CG of the
  non-purified system is also contained in some guild of the purified
  system.
\end{lemma}
\begin{proof}
  Observe that the $F_i \in \CF_i$ which $p_i$ removes cannot be in $\CT$,
  because otherwise it would be possible to cover all $\CP$ with sets in
  $\CT$; but this is not possible by the assumption $Q^3(\CT)$. This
  implies that the failure of $F_i$ cannot lead to the existence of a
  guild, and can be removed from $\CF_i$.  On the other hand, subsets of
  $F_i$ can possibly be in \CT, and $p_i$ keeps those subsets in $\CF_i$.
\end{proof}

Observe that the purification procedure is deterministic and uses
  information that is available to every process in the system:
  evaluating the $B^3$-condition, for example, already
  assumes that every process in the system knows the asymmetric fail-prone
  systems of the others and that Byzantine processes do not lie about
  their assumptions.
  
\begin{construction}[Composition of ABQS]
  \label{def:comp_asym}
  Let $\CP_1 = \{p_1, \ldots, p_{m + k}\}$ and
  $\CP_2 = \{p_{m+1}, \ldots, p_{n} \}$ be two sets of processes, with
  processes $p_{m + 1}, \ldots, p_{m+k}$ in common.  Let $\BQ_1$ be an ABQS
  on processes~$\CP_1$ with asymmetric fail-prone
  system~$\BF_1 = \{\CF^{(1)}_1, \ldots, \CF^{(1)}_{m + k}\}$, and $\BQ_2$
  an ABQS on processes~$\CP_2$ with asymmetric fail-prone
  system~$\BF_2 = \{\CF^{(2)}_{m+1}, \ldots, \CF^{(2)}_{n} \}$, where
  $\BF_1$ and $\BF_2$ are purified.  Moreover, let $\CT_1$ and $\CT_2$ be
  the tolerated systems of the two ABQS, respectively.  The \emph{composite
    fail-prone system} $\BF_3$ on processes $\CP_3 = \CP_1 \cup \CP_2$ is
\begin{equation*}
  \BF_3 = [ \CF^{(1)}_{1} \otimes \CT_2, \ldots, \CF^{(1)}_{m} \otimes
    \CT_2, \CF^{(1)}_{m+1} \otimes \CF^{(2)}_{m+1}, \ldots,
   \CF^{(1)}_{m+k}\otimes \CF^{(2)}_{m+k}, \CF^{(2)}_{m+k+1} \otimes \CT_1, \ldots,
    \CF^{(2)}_{n} \otimes \CT_1 ].
\end{equation*}      
  and the \emph{composite ABQS} $\BQ_3$ is any asymmetric quorum system for
  $\BF_3$.
\end{construction}

\begin{lemma}\label{lem:comp_asym}
  The composed fail-prone system $\BF_3$ resulting from Construction~\ref{def:comp_asym} satisfies the $B^3$-condition.
\end{lemma}

\begin{proof}
Towards a contradiction, let us assume that the $B^3$-condition does not hold on $\BF_3$. 
This means there exist processes $p_i$ and $p_j$ and fail-prone sets $ F_i \in \CF^{(3)}_i$, $F_j \in \CF^{(3)}_j$, and $F_{ij} \in {\CF^{(3)}_i}^* \cap {\CF^{(3)}_j}^*$ such that $\CP_3 = F_i \cup F_j \cup F_{ij}$. 
In the following we consider the restriction of $F_i, F_j$, and $F_{ij}$ to $\CP_1$, i.e., $F_i|_{\CP_1}, F_j|_{\CP_1}$, and $F_{ij}|_{\CP_1}$, respectively. 
We distinguish two cases for $p_i$ and $p_j$.
First, consider the case where $p_i$ and $p_j$ belong to different sets of processes and let, w.l.o.g., $p_i \in \CP_1 \setminus \CP_2$ and $p_j \in \CP_2 \setminus \CP_1$.
By the definition of the $\otimes$ operator, and with an argument similar to what we used in the proof of Lemma~\ref{lem:comp_sym_correct}, we get that $F_i|_{\CP_1} \in {\CF^{(1)}_i}^*$, that $F_j|_{\CP_1} \in {\CT_{1}}^*$, that $F_{ij}|_{\CP_1}$ is a common subset of $\CF^{(1)*}_i$ and $\CT_1^*$, and that their union covers $\CP_1$.
This is a contradiction because $\BF_1$ is purified.
Second, consider the case where at least one of $p_i$ and $p_j$ belongs to both $\CP_1$ and $\CP_2$,
and let, w.l.o.g., $p_i \in \CP_1, p_j \in \CP_1 \cap \CP_2$.
(If $p_i \in \CP_1 \cap \CP_2$ the same reasoning can be applied by projecting in $\CP_2$.)
For this case, we observe that
$F_i|_{\CP_1} \in {\CF^{(1)}_i}^*$, $F_j|_{\CP_1} \in {\CF^{(1)}_j}^*$, and that $F_{ij}|_{\CP_1}$ is a common subset of 
a fail-prone set in $\CF^{(1)*}_i$ and a fail-prone set in $\CF^{(1)*}_j$.
This contradicts the assumption that $B^3(\BF_1)$.
\end{proof}

\begin{remark}
Given an ABQS $\BQ_1$ for an asymmetric fail-prone system $\BF_1$ on processes $\CP_1$,
and an ABQS $\BQ_2$ for $\BF_2$ on $\CP_2$, 
and assuming that the processes of each BQS make no assumptions about processes in the other, a composition of the two systems
is only possible if the corresponding tolerated systems $\CT_1$ and $\CT_2$ both satisfy the $Q^3$-condition. 
This is because the processes of the first ABQS (and vice versa) are only external observers for the second system, and therefore only assess it through its tolerated system. 
Processes in $\CP_1$ want to make sure that whenever the second system is able to make progresses (that is, for every $T \in \CT_2$ that leads to a guild $\CG$), they will also be able to make progress.
To achieve this, they must consider all the $T \in \CT_2$ as a possible actual failed set. However, because the processes of the first system do not assume anything about the second system, the only way to achieve this is to include all the $T \in \CT_2$ in their fail-prone sets. 
This leads to an ABQS if and only if the $Q^3$-condition holds in the second system (and vice versa).
\end{remark}

Lemma~\ref{lem:comp_asym} and Lemma~\ref{lem:canon-asym} together imply the existence of an ABQS for $\BF_3$ as defined in Construction~\ref{def:comp_asym}. This is the asymmetric canonical quorum system $\BQ_3 = \overline{\BF}_3$.

For instance, let us consider two ABQS $\BQ_1$ and $\BQ_2$ on processes $\CP_1=\{p_1,\ldots,p_m\}$ and $\CP_2=\{p_{m+1},\ldots,p_n\}$ with asymmetric fail-prone systems $\BF_1$ and $\BF_2$, respectively, such that $\CP_1 \cap \CP_2 = \emptyset$. Then, the asymmetric canonical quorum system for $\BF_3$ is
\[ \BQ_3 = [\CQ_1 \cup \overline{\CT}_2,\ldots,\CQ_m \cup \overline{\CT}_2,\CQ_{m+1} \cup \overline{\CT}_1,\ldots,\CQ_n \cup \overline{\CT}_1],\]
where $\CQ_i = \overline{\CF}_i$, $\CQ_i \cup \overline{\CT}_j = \{Q_k \cup \CG_l~|~Q_k \in \CQ_i \land \CG_l \in \overline{\CT}_j\}$ and $\CG_i$ is a guild for a tolerated set in $\CT_j$. Notice that, by definition, $\overline{\CT}$ contains all the guilds that can be obtained within an ABQS. 

As a short proof of  why $\BQ_3$ is the canonical asymmetric quorum system of $\BF_3$, we observe that, by assuming $\CP_1 \cap \CP_2 = \emptyset$, the asymmetric fail-prone system $\BF_3$ in Construction~\ref{def:comp_asym} reduces to 
\[ \BF_3 = [\CF_1 \cup {\CT}_2,\ldots,\CF_m \cup {\CT}_2,\CF_{m+1} \cup {\CT}_1,\ldots,\CF_n \cup {\CT}_1],
\]
where $\CF_i \cup {\CT}_j = \{F_k \cup T_l~|~F_k \in \CF_i \land T_l \in {\CT}_j\}$. If we consider the bijective complement of $\CF_i \cup {\CT}_j$ this is made by all the sets of the form $\overline{F_k \cup T_l}$ in $\CP_3 = \CP_1 \cup \CP_2.$  Then, $\overline{F_k \cup T_l } = \overline{F}_k \cap \overline{T}_l = (Q_k \cup \CP_2) \cap (\CG_l \cup \CP_1)$ where $Q_k = \overline{F}_k$ in $\CP_1$. Finally, $(Q_k \cup \CP_2) \cap (\CG_l \cup \CP_1) = (Q_k \cap \CG_l) \cup (Q_k \cap \CP_1) \cup (\CP_2 \cap \CG_l) \cup (\CP_2 \cap \CP_1)$. Observe that, by assumption on the sets of processes, it follows that $(\CP_2 \cap \CP_1) = \emptyset$ and $(Q_k \cap \CG_l) = \emptyset.$ So, $\overline{F_k \cup T_l } = (Q_k \cap \CP_1) \cup (\CP_2 \cap \CG_l) = Q_k \cup \CG_l$.

\subsection{Composition in practice}

We now sketch a protocol that can be used by two (possibly disjoint) sets of  processes $\CP_1$ and $\CP_2$ that form two asymmetric Byzantine quorum systems $\BQ_1$ and $\BQ_2$ with asymmetric fail-prone systems $\BF_1$ and $\BF_2$, respectively.
We assume that processes in $\CP_1$ and $\CP_2$ are running two different instances of the same Byzantine consensus protocol (i.e., providing total-order broadcast)
and that $\BF_1$ and $\BF_2$ are publicly known.

The composition can be initiated by any process~$p_i$ in $\CP_1$.
To that end, process~$p_i$, acting as a client for $\BQ_1$, sends a \emph{composition-request} message to every process in $\CP_2$. Upon receiving this request, processes in $\CP_2$ start a round of Byzantine consensus:
if a sufficient number of processes votes for the composition, it will be agreed.
Assume the protocol instance run by $\BQ_2$  has a history of delivered messages $\CH_2$ at this point.
Then, upon deciding, processes in $\CP_2$ send a \emph{composition-response} message, which includes $\CH_2$, back to $\CP_1$. 

The rest of the protocol is symmetric to the first part; any process in $\CP_1$ that receives the same composition response from a guild of $\CP_2$ participates in a round of Byzantine consensus, this time within $\CP_1$. This results in $\CP_1$ sending a \emph{composition-acknowledgment} message to $\CP_2$, which now includes $\CH_1$,  the history of delivered messages in the instance run by $\CP_1$. The histories $\CH_1$ and $\CH_2$ can be used by the composed system to calculate the initial state of the new protocol instance, presumably using a generic \emph{merge function}.

The composition-acknowledgment message signals the start a new protocol instance run by $\CP_1 \cup \CP_2$.
From this point on, processes use the composed fail-prone and quorum systems. Since $\BF_2$ is known,
processes in $\CP_1$ can calculate both the tolerated system $\CT_2$ of $\BQ_2$ (in the simplest case by trying all possible failures of $\CP_2$) and the purified version of $\BF_2$, and vice versa for processes in $\CP_2$.
Should the fail-prone systems not be public, the processes could send them in the composition messages; however, privacy aspects are beyond the focus of this work.

\section{Conclusions}
\label{sec:conclusions}

Our work shows how trust assumptions of (possibly disjoint) systems can composed deterministically, such that groups of strangers may join each other and collaborate under a composed trust assumption with appealing properties. We present composition rules that work in both symmetric and asymmetric-trust models. Moreover, we overcome existing impossibility results for consensus among disjoint personal Byzantine quorum systems systems from the literature~\cite{DBLP:conf/wdag/LosaGM19}; given two systems that can reach consensus on their own, our composition method results in a system that achieves consensus. 
As intermediate results we define the tolerated system of an ABQS, which reflects the overall resilience of the ABQS, and present a purification procedure, which aligns the expectations of a process with the realistic capabilities of an ABQS. We expect these contributions to be of independent interest towards a deeper understanding and practical adoption of subjective decentralized trust.

\section*{Acknowledgments}

This work has been funded in part by the Swiss National Science Foundation
(SNSF) under grant agreement Nr\@.~200021\_188443 (Advanced Consensus
Protocols).

\bibliography{references, dblpbibtex}
\bibliographystyle{ieeesort}

\end{document}